# Dynamic Correlation Length Scales under Isochronal Conditions

R. Casalini, D. Fragiadakis, and C.M. Roland*

Naval Research Laboratory, Chemistry Division, Code 6120, Washington DC 20375-5342

*roland@nrl.navy.mil; 202-767-1719

**ABSTRACT**

The origin of the dramatic changes in the behavior of liquids as they approach their vitreous state – increases of many orders of magnitude in transport properties and dynamic time scales – is a major unsolved problem in condensed matter. These changes are accompanied by greater dynamic heterogeneity, which refers to both spatial variation and spatial correlation of molecular mobilities. The question is whether the changing dynamics is coupled to this heterogeneity; that is, does the latter cause the former? To address this we carried out the first nonlinear dielectric experiments at elevated hydrostatic pressures on two liquids, to measure the third-order harmonic component of their susceptibilities. We extract from this the number of dynamically correlated molecules for various state points, and find that the dynamic correlation volume for non-associated liquids depends primarily on the relaxation time, sensibly independent of temperature and pressure. We support this result by molecular dynamic simulations showing that the maximum in the four-point dynamic susceptibility of density fluctuations varies less than 10% for molecules that do not form hydrogen bonds. Our findings are consistent with dynamic heterogeneity serving as the principal control parameter for the slowing down of molecular motions in supercooled materials.

**SIGNIFICANCE**

On cooling or compressing a liquid, molecular motions become slower, effectively ceasing as the material becomes a glass. Coincident with this is development of reciprocity — neighboring species must make adjustments in order for a given molecule to move. Conceptually, slower motions and more cooperative movements would seem to be connected, as theoretical models often assume; however, experimental verification is lacking. We carried out nonlinear dielectric spectroscopy, which enables both the timescale and degree of cooperativity of molecular motions to be quantified. By obtaining results at elevated pressures and various temperatures, we determined that a given relaxation time is associated with a given degree of cooperativity, supporting the idea that cooperativity underlies the transition of a liquid to the glassy state.

The dynamics of liquids approaching their vitreous state exhibits interesting effects, the most prominent being spectacular changes in viscosity and relaxation times: these quantities may increase several orders of magnitude for a few degrees cooling. Eventually the response becomes so sluggish that the supercooled liquid behaves as a solid over laboratory time scales, and the material is now referred to as a glass. This super-Arrhenius slowing of molecular motions (i.e., the logarithm of the relaxation times increasing faster than linearly with reciprocal temperature) is concurrent with the growth of space-time correlations [1,2,3], as motion of a molecule increasingly requires adjustments of others. Inherent to spatial correlation of the molecular mobilities is dynamic heterogeneity, the spatial variation of the dynamics reflected in non-exponentiality of the relevant time-correlation function. The growth of transient order manifested as dynamic correlations is commensurate with the size of these dynamic heterogeneities, the latter having been measured experimentally [4,5]. Relatedly, computer simulations of model liquids have found that for a given material, a correlation exists between the dynamic correlation volume and the breadth of the relaxation dispersion [6].

Non-associated liquids, that is, liquids that lack hydrogen bonding, complex formation, etc., exhibit a property referred to as isochronal superpositioning, whereby the shape (breadth) of the relaxation function depends only on the relaxation time, $\tau$ [7,8]. Since the relaxation function reflects the distribution of molecular relaxation times (whose average is the $\tau$ measured experimentally), isochronal superpositioning implies that dynamic heterogeneity is intimately connected to $\tau$. Of interest herein is whether dynamic correlations and their growth might be the cause of, rather than only accompanying, the slowing down of the dynamics in the supercooled regime. The spatial extent of dynamic correlations exhibits power-law dependences on the relaxation time at constant (low) pressure, with the cooperativity extending to several intermolecular distances at the transition to the glassy state [1]. Such results suggest that dynamic correlations may serve as the control parameter governing vitrification [9,10,11,12,13]. However, to test this idea requires determination of the dynamic length scale in a liquid for different thermodynamic conditions having equivalent molecular mobility. Whether the dynamic correlation volume is invariant for state points with constant $\tau$ is a key to solving the glass transition problem [14].

**Quantifying dynamic correlation**

A number of indirect methods have been proposed to estimate dynamic correlation volumes [15,16,17,18,19], but precise determinations can be obtained from high-order correlation functions [1,20,21,22,23], such as the four-point dynamic susceptibility of the density fluctuations

$$\chi_4(t) = \int \left\langle \rho(r_1,0)\rho(r_1+r_2,0)\rho(r_1,t)\rho(r_1+r_2,t)\right\rangle_{r1} dr_2 \qquad [1]$$

$\chi_4(t)$ exhibits a maximum at a time $t \sim \tau$, with the height of the maximum proportional to the number of molecules dynamically correlated over this time scale: $N_c$=max{$\chi_4$}. At longer times $\chi_4$ decays to zero since there is no long-range, persistent order in an amorphous liquid. $\chi_4$ can be calculated in computer simulations, although results for model glass formers are mixed. Karmakar et al. [24] found that $\tau$ and the dynamic length scale have different dependences on the system size used in the simulation, inconsistent with the two quantities being coupled. On the other hand, $\chi_4$ calculated in the NVT ensemble for a Lennard-Jones model liquid [25] and the static length scale determined from point-to-set correlation functions (which measure the spatial extent of boundary effects) [26] were found to be correlated with the relaxation times.

Experimental determination of $\chi_4$ for real materials is problematic, requiring the use of approximations. Berthier et al. [11] expressed eq. [1] in terms of its various contributions, deriving two equations

$$\chi_4(t) = \frac{kT^2}{c_p}\left[\left.\frac{\partial C(t)}{\partial T}\right|_P\right]^2 + \chi_4^{NPH}(t) \qquad [2]$$

and

$$\chi_4(t) = \frac{kT^2}{c_V}\left[\left.\frac{\partial C(t)}{\partial T}\right|_\rho\right]^2 + \rho^3 kT\kappa_T\left[\left.\frac{\partial C(t)}{\partial \rho}\right|_T\right]^2 + \chi_4^{NVE}(t) \qquad [3]$$

In these equations $k$ is the Boltzmann constant, $c_P$ and $c_V$ are the respective isobaric and isochoric heat capacities, and $\kappa_T$ is the isothermal compressibility. The last term in either eqs. [2] or [3] is not available from measurements, but the insight of Berthier et al. [11] was the notion that both may be sufficiently small near $T_g$ that they can be neglected, with the dynamic susceptibility then approximated as

$$\chi_4(t) \approx \frac{kT^2}{c_p}\left[\left.\frac{\partial C(t)}{\partial T}\right|_P\right]^2 \qquad [4]$$

and

$$\chi_4(t) \approx \frac{kT^2}{c_V}\left[\left.\frac{\partial C(t)}{\partial T}\right|_\rho\right]^2 + \rho^3 kT\kappa_T\left[\left.\frac{\partial C(t)}{\partial \rho}\right|_T\right]^2 \qquad [5]$$

These expressions, both underestimates of $\chi_4(t)$, should be nearly equal.

Eqs. [4] and [5] have been applied to many materials [27], but of interest herein are measurements extending to high pressures, which enable assessment of the relationship between $N_c$ and $\tau$. Analyses using eq.[4] of four liquids, salol, polychlorinated biphenyl, propylene carbonate (PC), and a mixture of *o*-terphyenyl and *o*-phenyl phenol [28], found that state points having different temperatures and pressures but the same relaxation time had the same correlation volume; that is, $N_c$ is uniquely defined (to within ~10%) by $\tau$, or vice versa. Subsequently, from a similar analysis on dibutylphthalate [29] it was reported that at conditions of constant $\tau$, $N_c$ increased with pressure. A third study [30] used both eq. [4] and [5] and determined that $N_c$ for *o*-terphenyl, glibenclamide, and phenylphthaleindimethylether decreased with increasing $P$ at fixed $\tau$. Thus, three studies on 8 different liquids concluded that under isochronal conditions the dynamic correlation volume was constant [28], increased [29], or decreased [30] with increasing pressure and temperature.

These discordant results are not because the behavior of $N_c$ is material specific, but rather there are both fundamental and technical issues with the calculation of eqs. [4] and [5]. In simulations $\chi_4$ has a dependence both on the dynamics (Newtonian vs. Brownian) and the statistical ensemble (NVT, NPT, etc.) [31,32]. Moreover, it is unclear whether the heat capacity, the excess heat capacity (which isolates the configurational part relevant to structural relaxation), or the change in $c_p$ or $c_V$ at $T_g$ should be used in applying eqs. [2] – [5]; furthermore, values for the glassy or non-configurational part of the heat capacity are rarely available at high pressures. The approximation equations also involve derivatives of interpolated data, which can introduce uncertainties into the calculation of $N_c$. Note that eqs. [2] and [3] are equivalent, and their respective approximations differ only in the neglected last term, presumed to be negligible. However, in ref. [30] $N_c$ from eqs. [4] and [5] differed by as much as 40%; that is, the difference between two small contributions was an appreciable amount of the total $\chi_4$. This indicates either a lack of precision in the calculation of the approximations to $\chi_4$, or the assumption that $\chi_4^{NPH}$ and $\chi_4^{NVE}$ are small near $T_g$ [33] is incorrect (as has been found in simulations [34]). Notwithstanding the source of the problems in applying eqs. [4] and [5], the obtained results for the behavior of $N_c$ under isochronal conditions are ambiguous, motivating the use of a different method.

Although the linear susceptibility does not detect the transient order associated with dynamic heterogeneity, it has been argued that higher order correlation functions should be manifested in higher order (non-linear) susceptibilities [35]. An example is the non-linear magnetic response of spin glasses, which can detect transitions evident otherwise only in four-point correlation functions [36]. From scaling arguments applied to mode coupling theory, the response of liquids to external perturbations has been shown to be increasingly nonlinear as the glass transition is approached [37]. The inference is that the non-linear susceptibility can be used to measure dynamic heterogeneity; specifically, the amplitude of the nonlinear dielectric susceptibility is proportional to $N_c$ [35]

$$N_c \propto |\chi_3| \frac{kT}{\varepsilon_0 a^3 (\Delta\chi_1)^2} \qquad [6]$$

where $\varepsilon_0$ the permittivity of free space, $a$ the molecular volume, $\Delta\chi_1$ the linear dielectric strength, and $|\chi_3|$ is the modulus of the third-order susceptibility corresponding to polarization cubic in the applied field. The derivation of eq. [6] is more intuitive than rigorous, and features in the nonlinear susceptibility interpreted in terms of dynamic heterogeneity can be obtained in models that lack dynamic heterogeneity [38,39]. Nevertheless, this equation has been applied by several groups to determine the variation of $N_c$ at ambient pressure with $T$ [40,41,42] and during aging [43]. The main result, that $N_c$ grows on cooling towards $T_g$ in accord with the behavior of the relaxation time, supports the identification of the peak in $|\chi_3|$ with the dynamic correlation volume.

**Nonlinear dielectric results**

We carried out non-linear dielectric measurements under high pressure on two liquids, propylene carbonate and propylene glycol (PG). PC is a non-associated liquid conforming to isochronal superpositioning [7,8], whereas PG is hydrogen-bonded and thus its relaxation spectrum is not constant for constant $\tau$ [44]. In Figure 1 are representative $|\chi_3|$ spectra obtained at various pressures. For both liquids there is an increase in the peak intensity with decrease in peak frequency, consistent with growth

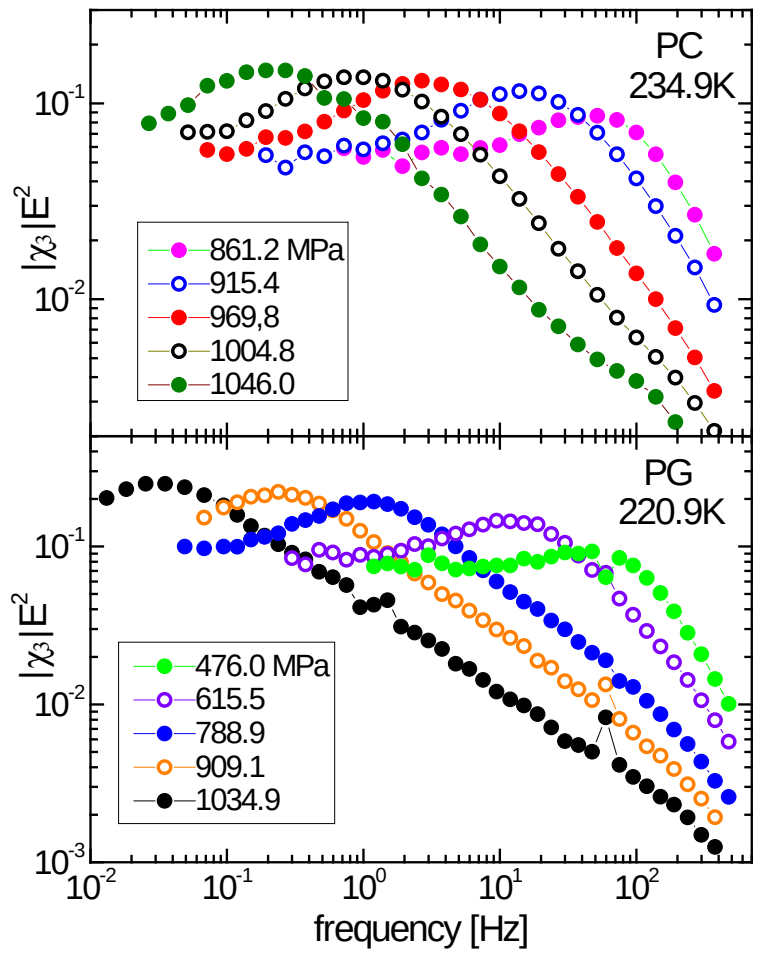

Figure 1. Representative third-order harmonic spectra of PC (top) and PG (bottom) at the indicated temperature and pressures, the latter increasing from right to left.

in the correlation volume as the relaxation time becomes longer.

To quantify dynamic heterogeneity requires that the contribution to $|\chi_3|$ from saturation of the dipole orientation be removed from the spectra. This saturation effect can be calculated assuming independent, rigid dipoles [40,45]

$$\chi_3^{sat}(\omega)=\frac{-3\varepsilon_0 a^3(\Delta\chi_1)^2}{5kT}\int_0^\infty g_{HN}\frac{3-17\omega^2\tau_0^2+i\omega\tau_0\left(14-6\omega^2\tau_0^2\right)}{\left(1+\omega^2\tau_0^2\right)\left(9+4\omega^2\tau_0^2\right)\left(1+9\omega^2\tau_0^2\right)}d\tau \quad [7]$$

The distribution of relaxation times (heterogeneous dynamics) are described using the Havriliak-Negami function [46]

$$g_{HN}(\ln\tau)=\frac{(\tau/\tau_0)^{\alpha\beta}\sin\beta\theta}{\pi\left((\tau/\tau_0)^{2\alpha}+2(\tau/\tau_0)^{\alpha}\cos\pi\alpha+1\right)^{\beta/2}} \quad [8]$$

where

$$\theta=\arctan\left(\frac{\sin\pi\alpha}{(\tau/\tau_0)^{\alpha}+\cos\pi\alpha}\right) \quad [9]$$

In the above $\alpha$ and $\beta$ are constants. We fit the linear dielectric loss peaks to eq.[8], then calculate the saturation effect using eq. [7]; typical results are shown in Figure 2. The difference spectrum can be used to extract $N_c$ via eq.[6].

As seen in Fig. 2, the saturation effect decays rapidly with frequency. This suggests a simpler method to avoid this interference, using a value of $|\chi_3|$ at higher frequencies than the peak. Brun et al. [41] have shown that $|\chi_3|$ at a frequency 2.5 times $f_{max}$, the frequency of the maximum in the linear dielectric loss, provides a measure of $N_c$ unaffected by dipole saturation. In the inset to Fig. 2 we show the results for $N_c$ of PC at ambient pressure obtained after correcting $|\chi_3|$ using eq. [7] and by taking the value of $|\chi_3|$ at 2.5$f_{max}$. Both values have similar temperature dependences, so that either method yields a quantity proportional to the correlation volume. Hereafter we report $|\chi_3(2.5f_{max})|$.

In Figure 3 are $N_c$ for the two liquids plotted as a function of the linear relaxation frequency. The data for PC show the two regimes expected for dynamic correlations – power-law dependences with a steeper slope at higher frequencies [11]. This supports the interpretation of the peak in the nonlinear susceptibility in terms of dynamic correlations. Within the experimental scatter (*ca.* 15%), the number of dynamically correlated molecules for PC depends mainly on the relaxation time; there is no systematic variation in $N_c$ with $T$ or $P$. We can also arrive at this result without any analysis, by simply comparing $|\chi_3|$ for two state points having the same peak frequency in their linear spectra. As seen in Figure 4, these $|\chi_3|$ spectra for PC superpose; thus, the isochronal $N_c$ are essentially constant.

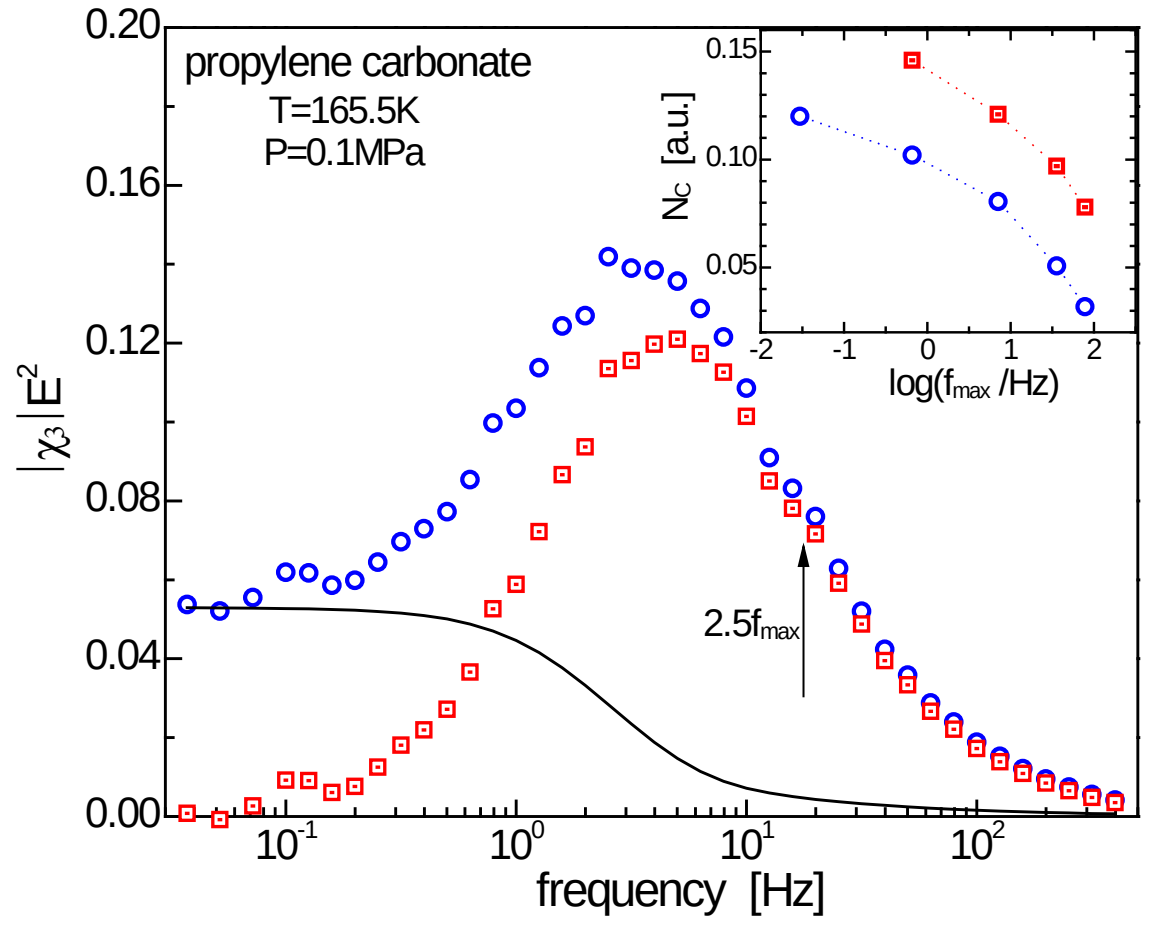

Figure 2. Third-order spectrum of PC as measured (circles) and after subtraction of the saturation effect (squares), the latter indicated by the solid line. Inset shows the magnitude at $2.5f_{max}$ in the uncorrected spectrum (circles) and at the peak of the corrected spectrum (squares).

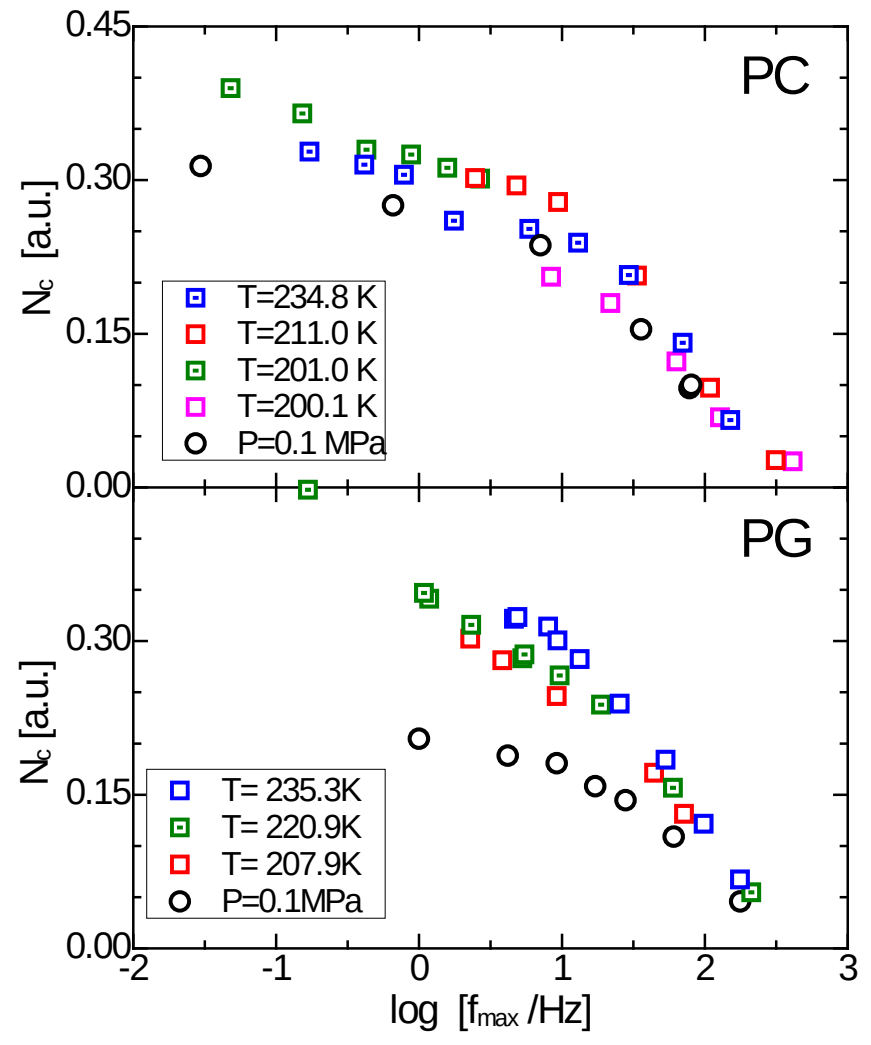

Figure 3. Number of dynamically correlated molecules (arbitrary units) for PC (top) and PG (bottom) as a function of the frequency of the loss peak in the linear spectrum. The axes scales are the same for both panels.

PG is an associated liquid and therefore is expected to behave differently. The results in Fig. 3 bear this out: there are substantial variations (>50%) in $N_c$ for a given $\tau$. The data indicate a systematic increase in the correlation volume with increasing temperature or pressure at constant $\tau$. The origin of this behavior is the change in H-bonding with thermodynamic conditions, whereby the liquid structure is not constant for isochronal conditions. Thus, neither the amplitude of $|\chi_3|$ nor the spectrum (Fig. 4) is invariant to changes in $T$ and $P$, even when $\tau$ remains constant.

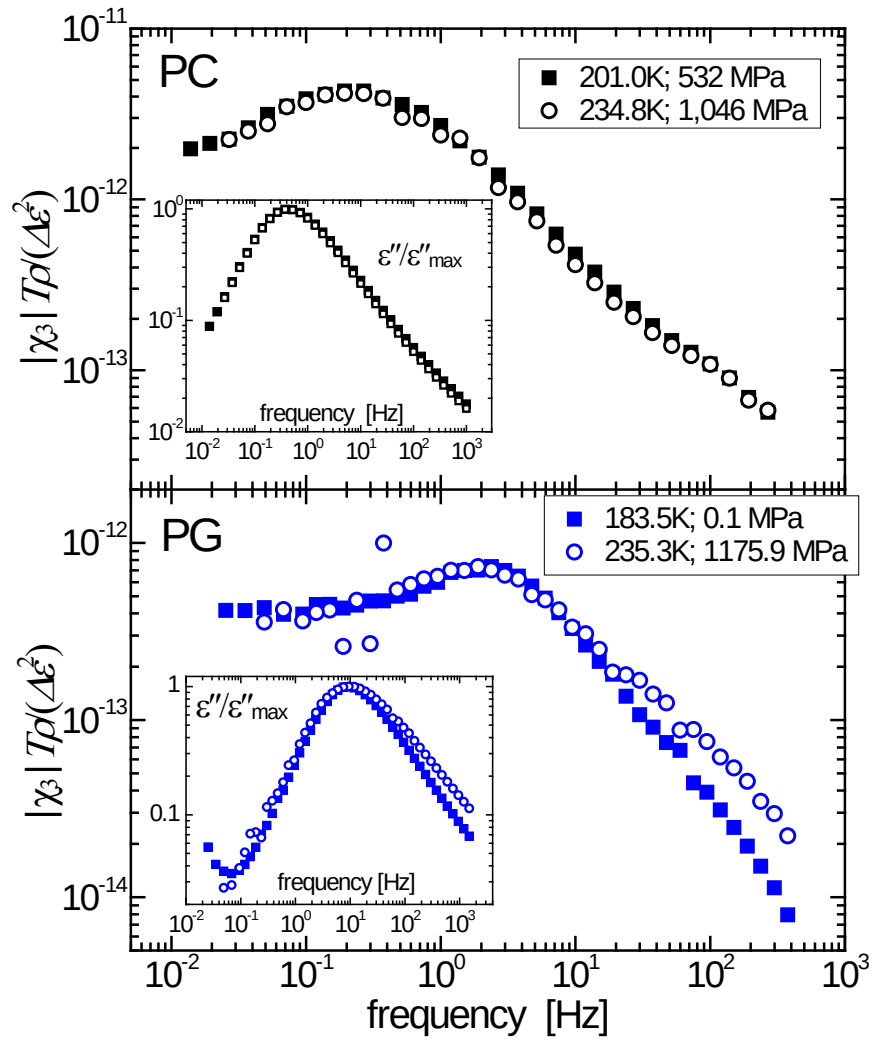

Figure 4. $|\chi_3|$ spectra for PC (top) and PG (bottom) measured at conditions corresponding to constant linear relaxation time. Only for the non-associated liquid is isochronal superpositioning observed, both for the nonlinear and linear (inset) spectra.

## Molecular dynamics simulations

As discussed above, the four-point dynamic susceptibility enables determination of $N_c$. Molecular dynamics (MD) simulations of Lennard-Jones particles found that the $N_c$ from $\chi_4$ was constant for constant $\tau$ [25]. This result is also predicted for simple liquids in the NVT ensemble [32], "simple" defined as liquids displaying certain properties, such as isochronal superpositioning of the relaxation function. Accordingly, to compare with the dielectric results herein, we carried out MD simulations on two rigid, polar united-atom structures, modelling a three-site hydrogen-bonded molecule and a two-site structure with the same dipole moment but no H atom (and thus non-associating). Details of the simulations can be found in ref. [47]; the difference herein is a larger system size (10,000-16,000 molecules).

To compare to the dielectric relaxation results, we calculated $\chi_4$ for dipole reorientation from the variance of the rotational autocorrelation function, $C_1(t)$ (rather than eq.[1], the 4-point correlation function for density fluctuations)

$$\chi_4^{NVT}(t) = N\left[\left\langle C_1(t)^2\right\rangle - \left\langle C_1(t)\right\rangle^2\right] \qquad [10]$$

The linear susceptibility

$$\chi(\omega) = \chi'(\omega) + i\chi''(\omega) = 1 + \mathrm{i}\,\omega\int_0^{\infty} e^{i\omega t}\phi(t)\,\mathrm{d}t \qquad [11]$$

was also computed to assess the susceptibility loss spectra at various state points with constant relaxation time, as done for the experimental dielectric spectra.

$\chi_4$ determined for the NVT ensemble for the non-associated liquid at four isochronal state points are shown in Figure 5. The peak heights, and thus $N_c$, are within 4% for density variations of as much as 16%. For the H-bonded structure (Figure 6), this variation under isochronal conditions is much larger, more than a factor of two. Similarly, there is a breakdown of isochronal superpositioning, as shown in the inset to Fig. 6. The very different behavior of the non-associated and the H-bonded materials, as summarized in Figure 7, corroborate the nonlinear dielectric determination that for propylene carbonate, but not for the H-bonded propylene glycol, $N_c$ is sensibly constant at constant $\tau$, its variation being a few percent. This behavior is consistent with isochronal superpositioning of the linear relaxation function for PC, but not for PG.

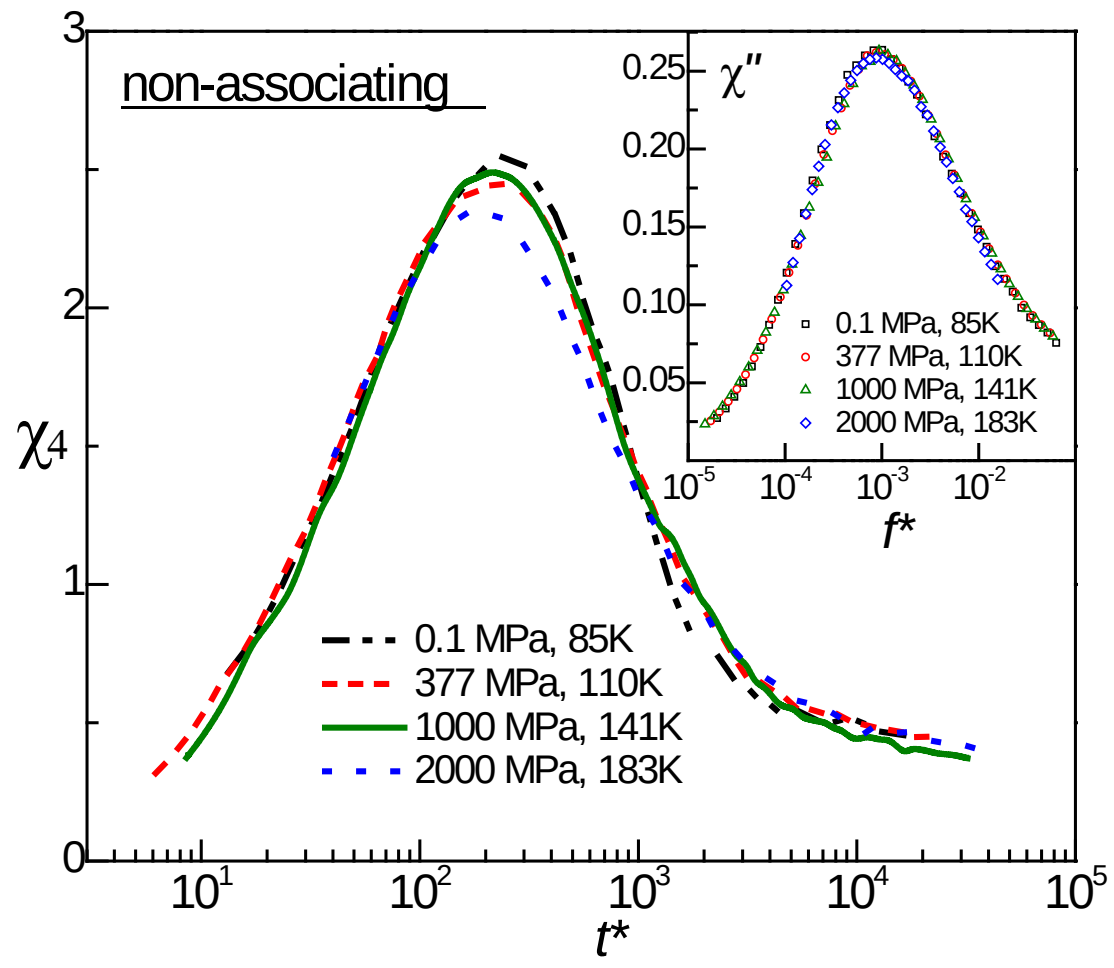

Figure 5. $\chi_4$ for the dipole autocorrelation function for a polar, non-associating liquid at three state points having equal reduced relaxation times ($t^*=\rho^{1/3}(kT)^{1/2}t$). The density change between the highest and lowest pressure data is 16%. The inset shows the linear susceptibility for the same state points versus reduced $1/t^*$.

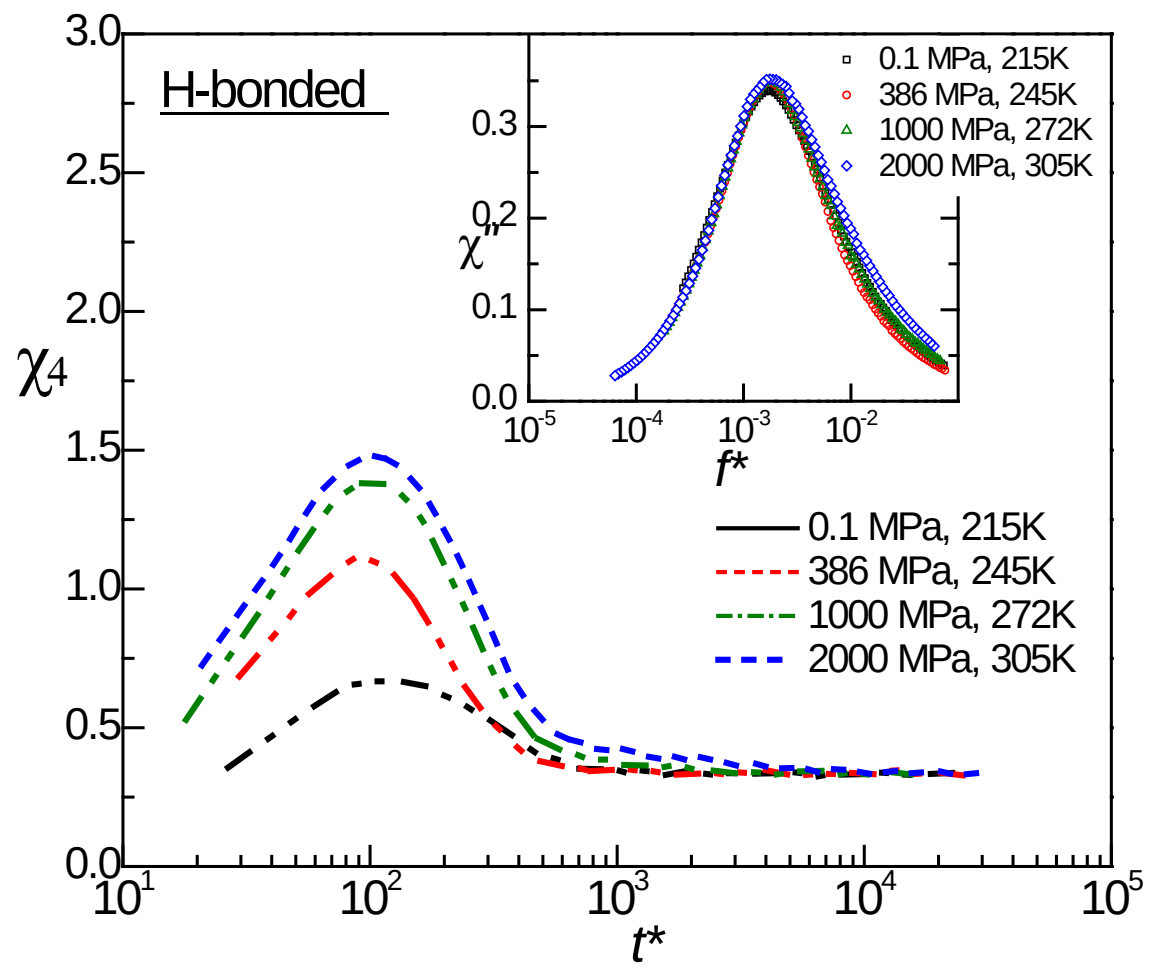

Figure 6. $\chi_4$ for the dipole autocorrelation function for a hydrogen-bonded liquid at four state points having equal reduced relaxation times. Inset shows departure of linear susceptibility from isochronal superpositioning.

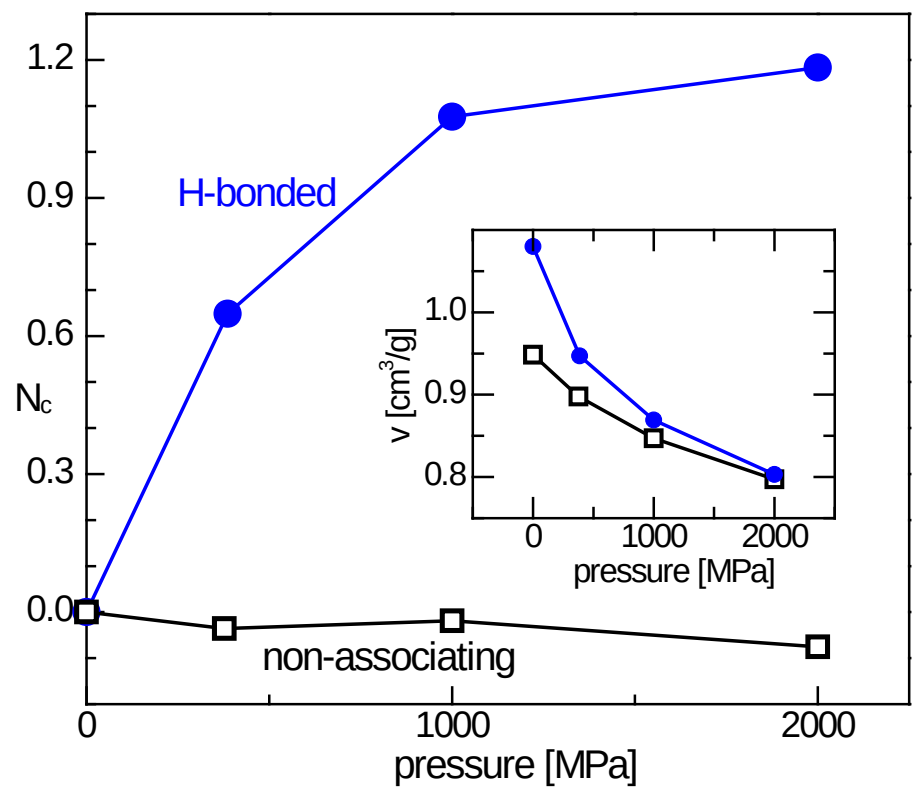

Figure 7. Pressure-variation of the maximum in $\chi_4$ at temperatures for which the reduced relaxation time $\tau^*$ is constant ($\tau^*$~$10^2$) for the non-associated liquid (open squares) and the H-bonded liquid (filled circles); the variation in the latter is substantially larger. Specific volume at same (isochronal) state points shown in inset.

## Summary

For non-associated liquids, the third-order dielectric susceptibility and four-point dynamic susceptibility from MD simulations both show that there is only a small (~10%) variation in $N_c$ for state points having significant differences in density but the same relaxation time. Together with prior work showing a correlation between spatial variation of molecular mobilities and $\tau$, this means that these two reflections of dynamic heterogeneity are fundamentally connected to the time scale of the dynamics time. This is important for development of a theory of the glass transition, since viable models that predict $\tau$ implicitly make simultaneous predictions for $N_c$ and the distribution of relaxation times.

**Materials and Methods**

Samples for nonlinear dielectric experiments were contained within Teflon gaskets, 10 – 25 $\mu$m diameter, and between polished steel parallel plates that served as the electrodes. This capacitor assembly was placed in a Manganin cell (Harwood Engineering) inside a Tenney environmental chamber. Pressure was applied via using an Enerpac pump in combination with a Harwood pressure intensifier; ?? was used as the hydraulic fluid. The voltage for the nonlinear experiments was as large as 200 V (rms), with a Novocontrol HVB 4000 dielectric analyzer used for th measurements.

The molecular volumes, *a(T,P)* in eq. [6], were obtained from the published equation of states for PC [49] and PG [[50]].

MD simulations were carried out using GROMACS [46]. For details of the system see ref. [48]. Larger system sizes were used herein (N=10000 for the H-bonded and N=16000 for the non-associated liquid).

**Acknowledgment.** This work was supported by the Office of Naval Research. We thank R. Richert for informative discussions on nonlinear dielectric spectroscopy.